# Multi-Agent Based Simulation for Investigating Electric Vehicle Adoption and Its Impacts on Electricity Distribution Grids and CO2 Emissions

Christensen, Kristoffer; Ma, Zheng Grace; Jørgensen, Bo Nørregaard

Go to publication entry in University of Southern Denmark's Research Portal

# Multi-Agent Based Simulation for Investigating Electric Vehicle Adoption and Its Impacts on Electricity Distribution Grids and CO2 Emissions

Kristoffer Christensen [1][0000-0003-2417-338X], Zheng Ma[1][0000-0002-9134-1032]

and Bo Nørregaard Jørgensen[1][0000-0001-5678-6602]

[1] SDU Center for Energy Informatics, Maersk Mc-Kinney Moeller Institute, The Faculty of Engineering, University of Southern Denmark, Odense, Denmark
kric@mmmi.sdu.dk, zma@mmmi.sdu.dk, bnj@mmmi.sdu.dk

**Abstract.** Electric vehicles are expected to significantly contribute to CO2-eq. emissions reduction, but the increasing number of EVs also introduces challenges to the energy system, and to what extent it contributes to achieving climate goals remains unknown. Static modeling and assumption-based simulations have been used for such investigation, but they cannot capture the realistic ecosystem dynamics. To fill the gap, this paper investigates the impacts of two adoption curves of private EVs on the electricity distribution grids and national climate goals. This paper develops a multi-agent based simulation with two adoption curves, the Traditional EV charging strategy, various EV models, driving patterns, and CO2-eq. emission data to capture the full ecosystem dynamics during a long-term period from 2020 to 2032. The Danish 2030 climate goal and a Danish distribution network with 126 residential consumers are chosen as the case study. The results show that both EV adoption curves of 1 million and 775k EVs by 2030 will not satisfy the Danish climate goal of reducing transport sector emissions by 30% by 2030. The results also show that the current residential electricity distribution grids cannot handle the load from increasing EVs. The first grid overload will occur in 2031 (around 16 and 24 months later for the 1 million and 775k EVs adopted by 2030) with a 67% share of EVs in the grid.

**Keywords:** electric vehicle, adoption curve, distribution grid, agent-based modeling, multi-agent systems, CO2 emissions

## 1    Introduction

The transportation sector contributes greatly to CO emissions. For instance, In 2019 cars in Denmark emitted 7.2 million tons of CO2-eq, and the total car fleet in Denmark in 2019 was 2.65 million [3], resulting in an average emission per car of 2.72 tons annually. The car fleet is estimated to increase to about 3.25 million in 2030 [2]. With no change in car emissions, this would result in a total car emission of 8.84 million tons by 2030.
The European Union's climate policies (which Denmark is obligated to), states among other things that emissions from buildings, agriculture, and transportation have to be reduced by 30% compared to 2005 levels [1]. In 2005 the CO2-eq. emissions from cars were around 7.1 million tons [1]. This corresponds to a maximum emission level of 4.97 million tons CO2-eq. from cars by 2030, in order to meet the climate goal. Assuming Electric Vehicles (EVs) have zero emissions, a minimum of 1.42 million of the car fleet in 2030 has to be EVs to meet the climate goal. However, EVs do indirectly have a CO2-eq. emission until the day the electricity grid consists of 100% green energy sources.

Furthermore, due to the regulations, electricity Distribution System Operators (DSOs) have no access to information on how many EVs, when, and how much EVs will charge in the distribution grids with the increasing number of EVs. Therefore, DSOs are unclear about the loading profile of the distribution grids in the future. Although many studies have investigated EV-caused overloads in distribution grids (e.g., [4, 5]), most studies are based on static modeling or assumption-based simulations without considering the real EV adoption curves. It causes a challenge for DSOs that it is unclear what the overloads in distribution grids will look like with the increasing number of EVs over the years in the future. However, the majority of the literature is based on static modeling or assumption-based simulations without considering the realistic EV adoption curves. It would result in high uncertainty for DSOs to conduct grid planning, especially congestion management.

Therefore, this paper aims to investigate how different adoption curves of private EVs will impact the national climate goals and the consequence to the electricity distribution grids. A multi-agent based simulation with several adoption curves, EV models, driving patterns, and CO2-eq. emission from EV charging is developed to capture the full ecosystem dynamics during a long-term period from 2020 to 2032 with a high resolution (hourly).

Furthermore, the adoption curves, EV models, and driving patterns are all based on national statistics and national market research to ensure the simulation can closely represent the reality to provide a clear and realistic load profile of distribution grids in the future due to EV adoption. The CO2-eq. emissions from consuming electricity depend on the electricity production mix in the grid, thus, the simulation imports and extrapolates hourly electricity consumption emission data. A distribution grid consisting of 126 residential consumers in the city of Strib, Denmark is selected as a case study to investigate the uncertainty of the increasing EVs' impact on the distribution grid with the electricity consumption data for 2019.

The paper outline starts with a background of the Danish electricity system and electric vehicle adoption curve; secondly, the methodology is introduced followed by the case study; thirdly, the scenario and experiment design is presented; lastly, the results are presented followed by the discussion and conclusion.

## 2  Literature review

The applications of Multi-Agent Systems (MASs) in the energy domain are increasing as the energy system complexity increases as a result of the green transition. [6] conduct a scoping review of the literature on ontology for MAS in the energy domain. [6] identify the energy domain applications for MAS to be within grid control, electricity markets, demand-side, and building systems. This paper's MAS should represent the energy ecosystem around EV home charging as proposed in [7]. The MAS representation of the ecosystem simulates the impact on the distribution system from the adoption of EVs over time.

The adoption of technology and innovation is a well-covered subject since the innovation adoption theory was first introduced in 1960 by Everett Rogers in his book called "Diffusion of Innovation Theory" [9]. 30 technology adoption theories have been identified [8], and essential elements from the theory are used, such as the S-shaped (logistic function) and the adoption rate curve. Several studies, e.g., [10-14], are conducted using agent-based modeling together with Rogers' adoption theory. Furthermore, [15] and [16] use MAS for investigating the impact of adopting EVs, but do not consider grid loading, tariff schemes, or charging algorithm adoption. [15] uses an ecosystem approach, but the ecosystem does not consider the business part of the ecosystem, hence not considering several flows (e.g., monetary flows). [15] focuses on spatial adoption and does also not consider the business ecosystem perspective. Moreover, [17] investigates the CO2 emissions from charging EVs and compares them with plug-in hybrid EVs and conventional vehicles. The paper applies a static approach using a fixed CO2 intensity of the electricity generation mix to calculate the emissions per km driven by the EV.

## 3  Methodology

This paper chooses MAS to investigate the ecosystem dynamics, stakeholders' behaviors, and the impacts of increasing EVs on an energy business ecosystem. MAS is chosen since it allows an assembly of several agents with either homogenous or heterogeneous architecture [19] and has been popularly applied to model and simulate complex systems.

The selection of the agent-based simulation tool applied in this paper is based on a comprehensive comparative literature survey of the state-of-the-art in software agent-based computing technology [20]. The survey addresses more than 80 software tools. Classifications are made considering the agent-based simulation tools' scope or application domain and the computational modeling strength against the model development effort. Furthermore, three criteria are defined for the evaluation and selection:

- High to extreme scale of computation modeling strength and simple to moderate model development effort
- The application domains are in a dynamic computational system, business, economics, planning & scheduling, enterprise, and organizational behavior
- Suitable for simulating energy business ecosystems

Among 80 software tools, AnyLogic seems to fit the purpose as AnyLogic's application domains cover power grids, business strategy & innovation analysis. Hence, AnyLogic is chosen as the simulation tool in this paper.

## 4  Multi-Agent Based Simulation Development

There are a vast number of potential scenarios for complex MASs with a wide variety of parameter inputs to investigate an ecosystem. However, in this paper, the focus is to enlighten impacts on the CO2-eq. emissions and distribution grid from increasing EVs. To do so, the ecosystem is translated into a MAS by identifying agents, interfaces, and communications. The identified MAS elements are programmed in the selected software tool AnyLogic.

The agents represent actors and objects in the ecosystem. The communication flows are between roles. Therefore, the roles are represented by Java interfaces (Anylogic is Java-based) holding the interactions related to the role. The agents implement the respective interfaces corresponding to the actor/object and the associated roles.

Fig. 1 shows a screenshot of the running simulation which shows an overview of the simulated ecosystem, which is based on the ecosystem presented in [7]. The states of the households and transformer loading are shown in real-time.

The relevant input data for this paper are the adoption curves, EV models, driving patterns, and CO2-eq. emission from EV charging.

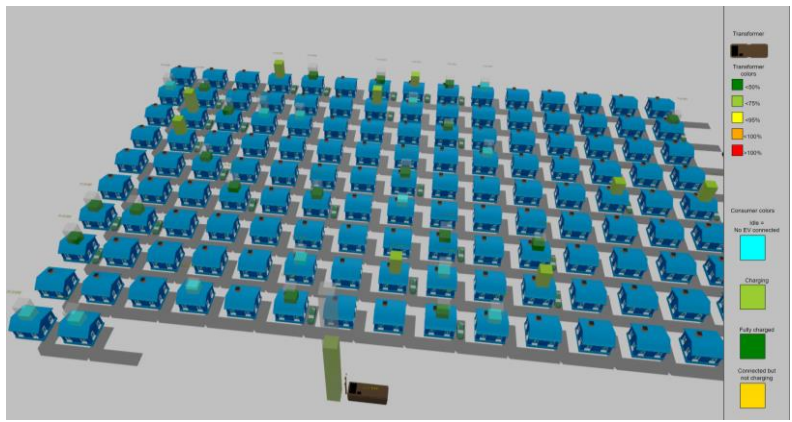

**Fig. 1.** Screenshot of the electric vehicle home charging multi-agent system.

### 4.1  Danish Electric Vehicle Adoption Curve

Based on Rogers adoption curves and the national statistics [18], the residential EV adoption in Denmark from 2011 to 2021 is shown as the black line in Fig. 2 which is close to a logistic growth with a 52.6% rate shown as the orange line in Fig. 2. The logistic function (Roger's S-curve) running through 124 EVs in 2011 and 16,687 in 2021 is identified using equation 1 (orange line in Fig. 2).

$$P(t) = \frac{A}{1+\left(\frac{A}{P(0)}-1\right)e^{-rt}} \quad (1)$$

Where $P(t)$ is the number of EVs to the time $t$. $A$ is Denmark's total number of vehicles in December 2021 of around 2.5 million cars for residents [18]. $P(0)$ is the initial value in January 2011 of 124 EVs, and $r$ is the growth rate in percent. The growth rate is identified to be 52.6% in order to match the EV adoption in 2011 and 2021. This results in a population of EVs of 1.3 million by 2030.

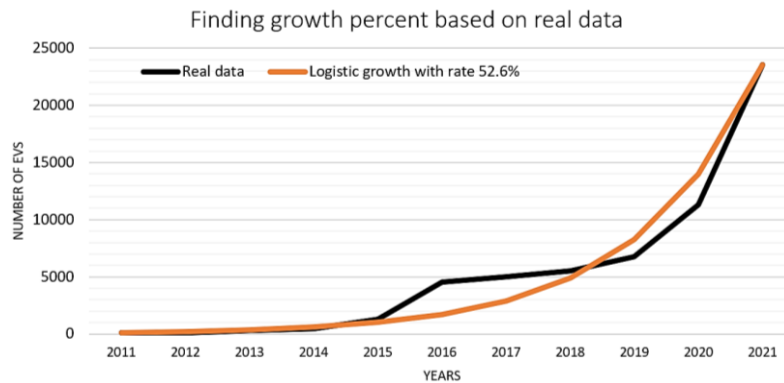

**Fig. 2.** Growth function estimation based on historical data (orange).

### 4.2 EV models

The EV models are selected as the five most sold EV types in 2019 in Denmark, and the probability of which EV is adopted is based on the purchase share of each type in 2019 (shown in Table 1) [21]. This general approach can be used for other markets by identifying the most popular EV models and their adoption shares.

**Table 1.** Top five most sold electric vehicles in Denmark in 2019.

| EV model | Capacity [kWh] | Mileage [kWh/km] | Maximum charging power [kW] | Percentage of purchased EVs in 2019 [%] [21] |
|---|---|---|---|---|
| Tesla Model 3 [22] | 50 | 0.151 | 11 | 40.5 |
| VW e-Golf [23] | 35.8 | 0.168 | 7.2 | 18.0 |
| Hyundai Kona [24] | 42 | 0.154 | 11 | 08.3 |
| Renault Zoe [25] | 44.1 | 0.161 | 22* | 06.0 |
| Nissan Leaf [26] | 40 | 0.164 | 3.68 | 05.8 |

*Charging power is limited to the maximum power that can be consumed by each household in Denmark. This limit is typically three phases of 25 Amps (corresponding to approximately 17.3 kW) [27].

### 4.3 Driving patterns

This research defines the EVs' departure and arrival times based on the significantly decreased and increased electricity consumption data in the mornings and evenings. A significant increase/decrease is defined as 80% above idle hours. Idle hours are calculated as the average load between hours 0 and 5 (at night). Those hours are assumed to represent the idle load of the consumer (consumption when residents are sleeping). When identifying a significant increase in load (indicating residents are awake) between 5 and 9 AM (default), the departure time is set within the hour, which has a load below the significant level (indicating residents have departed). The arrival time is within the hour when the load increases above the significant level between 2 and 10 PM (default). Fig. 3 illustrates how departure and arrival are chosen for a consumer.

Suppose no significant decrease or increase in consumption data is detected. The default time is set to a random hour from 5 to 9 AM and 2 to 10 PM for departure and arrival hours, respectively. The same approach applies to systems with similar working cultures, working from around 8 AM to 4 PM.

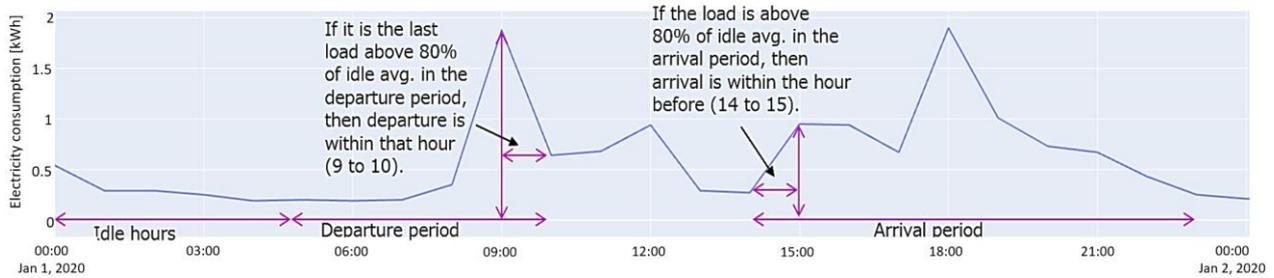

**Fig. 3.** Example of how the departure and arrival times are chosen.

The distribution of the driving distance per EV is adapted from [28]. The driving distance statistics in [28] are based on 100,000 interview data collected over 15 years for private Danish vehicle users. For simplicity, all EVs will be driven once a day.

### 4.4 EV charging CO2-eq. emission

The CO2-eq. emission from consuming electricity is imported from [29] for DK1 on a 5-minute resolution from 2017 to 2020 (4 years). The data is converted to hourly-based (hourly average) to match consumption data. The data are illustrated in Fig. 4.

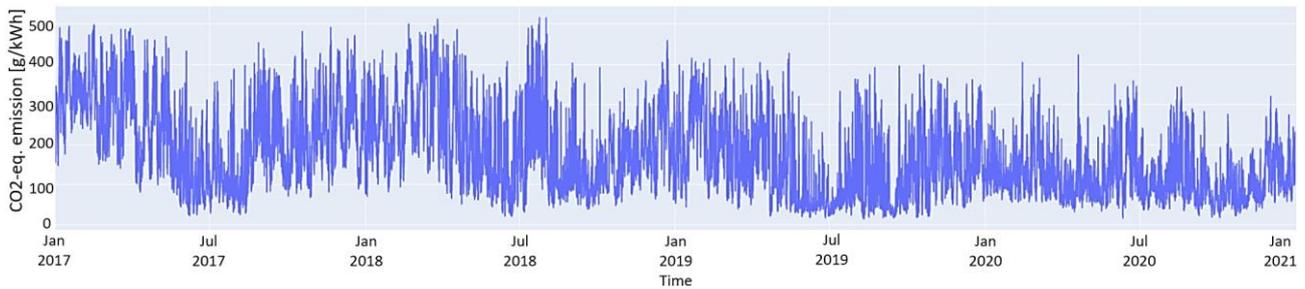

**Fig. 4.** Data on CO2-eq. emissions from consuming electricity, from 2017-2020.

The Danish Energy Agency [30] has extrapolated the CO2-eq. emission factor for electricity from 2020 to 2030. The extrapolation is shown in Fig. 5. An exponential regression is made to identify a function that describes the reduction of CO2-eq. emissions from electricity over the years without becoming negative. The Danish Energy Agency's extrapolation is considered a valid calculation used as the reduction factor in the implemented dataset.

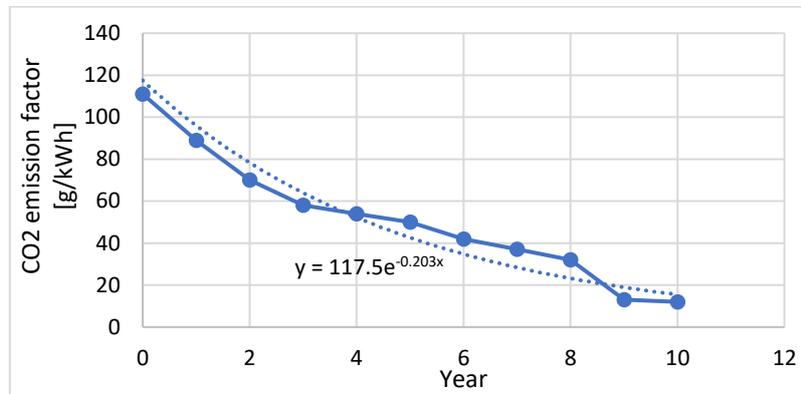

**Fig. 5.** CO2-eq. emission factor extrapolation over time.

From the exponential regression shown in Fig. 5, the emission factor can be derived by equation 2.

$$\text{Emission}_{new} = \text{Emission}_{data} \cdot e^{-0.203 \cdot \text{Year}} \qquad (2)$$

Where *Emission*$_{new}$ is the new calculated emission accounting for the reduction depending on the simulation time. *Year* is the simulation-years past. Fig. 10 shows the data after implementing the emission factor extrapolation during simulation time. The figure shows the emissions reductions over time as the electricity becomes greener, giving a more realistic result on the environmental impact from different scenarios.

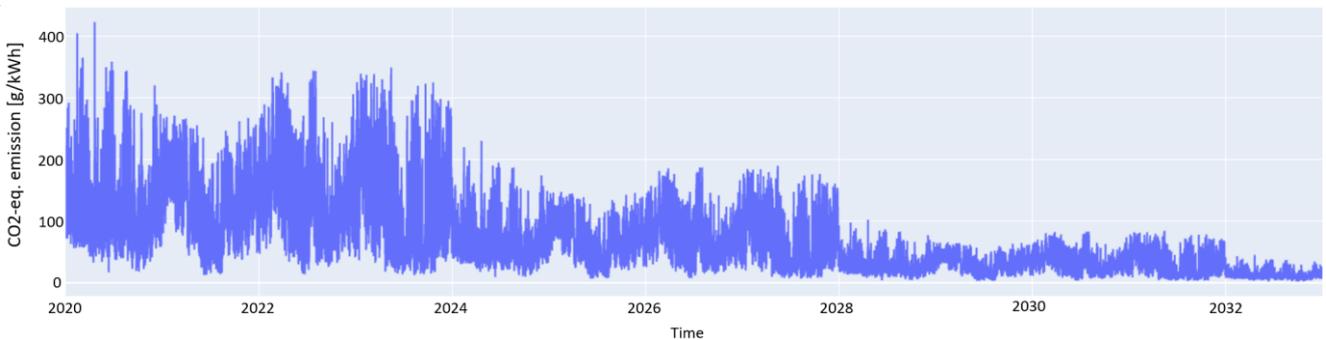

**Fig. 6.** Repeated data on CO2-eq. emissions from consuming electricity with the implementation of emission factor extrapolation (simulated).

Table 2 shows the results of interest based on the relevant stakeholders' (i.e., the DSO and households) perspective. The output is exported at the simulation end for analysis.

**Table 2.** Scenario outputs based on the relevant stakeholders.

| Stakeholder | Considered output |
|---|---|
| DSO | <ul><li>Load Factor (for the day with the first overload)</li><li>Coincidence factor (for the first year after overload)</li><li>Date and time for the first overload</li><li>Number of overloads first year after the first overload</li></ul> |
| EV users | <ul><li>Total average of all EV users' average CO2-eq. emissions.</li><li>Average annual CO2-eq. emissions from charging EVs (kg CO2-eq.).</li></ul> |

## 5 Case Study

The Danish electricity grid is divided into DK1 (Western Denmark - Jutland and Funen) and DK2 (Eastern Denmark - Zealand). The electricity grid is divided into generation, transmission, distribution, and consumption. This paper focuses on the Danish low voltage (400V) distribution grid. This paper's electricity system boundary starts from the 10kV/0.4kV transformer and ends at the residential consumers. The components between the transformer and consumer (i.e., cables, nodes, etc.) are not considered. A distribution grid consisting of 126 residential consumers in the city of Strib, Denmark, is selected as a case study to investigate the uncertainty of the increasing EVs' impact on the distribution grid.

The EV adoption for the area with 126 residents is shown in Fig. 7, which is an extrapolation from the logistic function identified in Fig. 2 (in Section 4.1 - Danish Electric Vehicle Adoption Curve). Consumption data for 126 residential consumers are provided by the Danish DSO TREFOR, which operates in the area of Strib.

The data for Strib is available from 2019 to now. However, since this paper uses consumption data to estimate driving behavior, the years with COVID-19 have been excluded, i.e., 2020 and 2021. Therefore, the data is used for 2019 alone. Furthermore, the data has been cleaned for households with EVs, PVs, heat pumps, electric heating, and missing data points. This ensures that the simulation has the actual number of consumer consumption patterns without any distributed energy resources, such as heat pumps. The consumption data is for 2019 and are shown in Fig. 8. The figure does not show the actual year on the Time-axes since they are produced from the simulation outputs and are repeated yearly.

The baseline scenario uses the case study's adoption curve shown in Fig. 7. The model does not apply the adoption curve directly but uses the yearly adoption as the average rate in a Poisson process [31]. This means that the EVs are adopted each year randomly at the given adoption rate, which on average, corresponds to the number of EVs adopted within that year. The reason is to take the uncertainty into account. However, for the scenario of various EV adoption curves, the curves are used directly for comparison purposes.

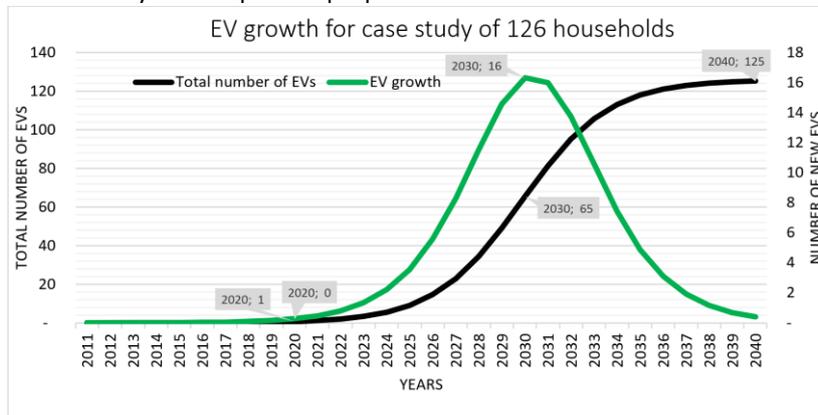

**Fig. 7.** Adoption curve (black) and adoption rate i.e., electric vehicles adopted per year (green).

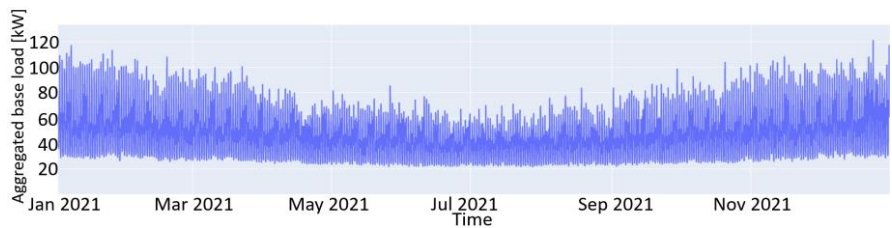

**Fig. 8.** Aggregated hourly consumption of the 126 consumer consumption data.

## 6 Scenario Design

Two scenarios are considered in this paper with two EV adoption curves of 775k and 1 million EVs in Denmark by 2030. The scenarios are identified based on the estimation and the political goal by [32], respectively. The two adoption curves are shown in Fig. 9 and Fig. 10, representing the two designed experiments. All simulation experiments start, by default, in 2020 and stop one year after experiencing the first overload. The results considered as key results for this paper are the date for the first overload, the frequency of overloads in the following year, and the average CO2-eq. emissions from charging the EVs. The average total CO2-eq. emissions are calculated for 2031, as this is relevant for the 2030 goal of reducing transport sector emissions by 30%. 2031 has been chosen to see if the goals by 2030 have been achieved in different scenarios.

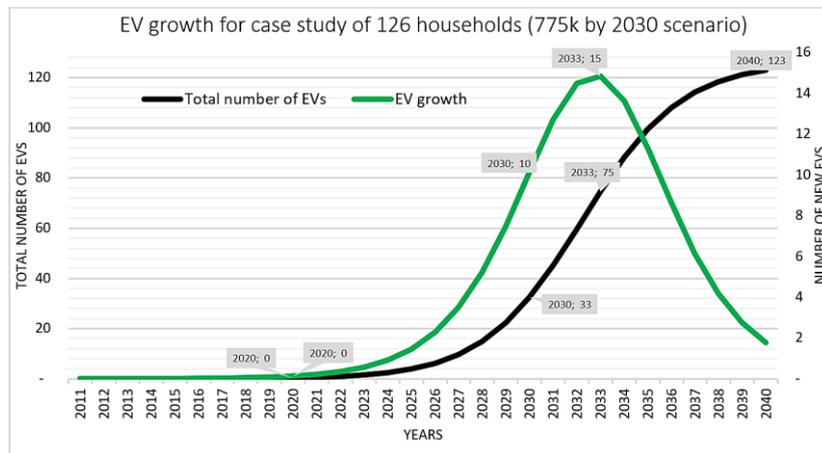

**Fig. 9.** 775k electric vehicles by 2030 adoption curve applied to the case study.

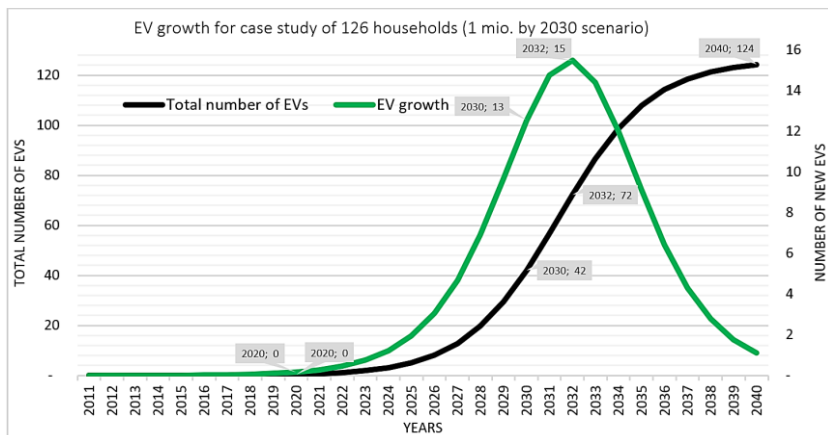

**Fig. 10.** 1 million electric vehicles by 2030 adoption curve applied to the case study.

## 7 Results

### 7.1 Baseline Scenario Results

The baseline scenario is simulated from 2020 to 2032 and simulates the Traditional charging strategy (i.e., charging immediately at arrival). The baseline scenario's EV adoption curve with the Poisson process from the simulation start (2020) to the last simulation year (2032) is shown to the left in Fig. 11. By the end of 2032, 98 EVs are adopted. The distribution of the EV models at the beginning of the last simulation year (January 1, 2032) is illustrated to the right in Fig. 11. The key results for the scenario are shown in Table 3.

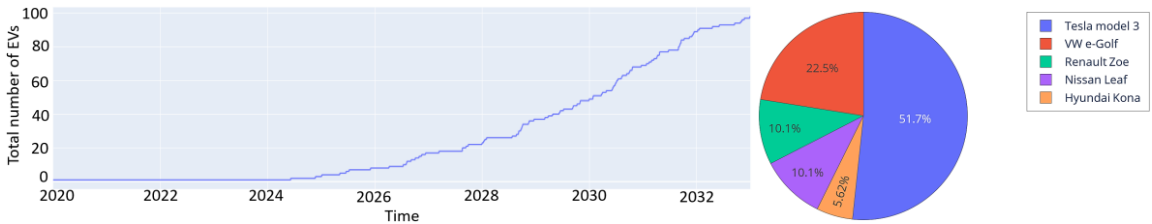

**Fig. 11.** The total number of adopted electric vehicles from the simulation start (2020) to the simulation end (2032) (left) and the distribution of electric vehicle models at the beginning of the last simulation year (January 1, 2032) (right).

**Table 3.** Baseline scenario key results.

| Date of the first overload | Overloads the year after the first overload | Days with overload | Total EVs when the first overload occurs | Avg. total CO2-eq. emissions in 2031 [kg] |
|---|---|---|---|---|
| Oct. 21, 2031, 4 PM | 60 | 41 | 85 | 116.37 |

Fig. 12 shows grid information for the day with the first grid overload (October 21, 2031). The first overload in the baseline scenario occurs on October 21, 2031, at 4 PM. A total of 85 EVs are adopted in the grid, with 49 simultaneous charging EVs. The first overload has a size of 1.44 kW above the grid capacity. At this overload, the total charging load is 324.92 kW. The orange line in Fig. 12 shows the total electricity prices and that there is one peak-price period during winter due to the DSO tariff. However, the EVs are not considering the electricity price when charging (Traditional charging strategy, i.e., charging immediately at arrival) and the result from this is 60 overloads (as shown in Fig. 13) the year after the first overload. Nine additional EVs are adopted that year.

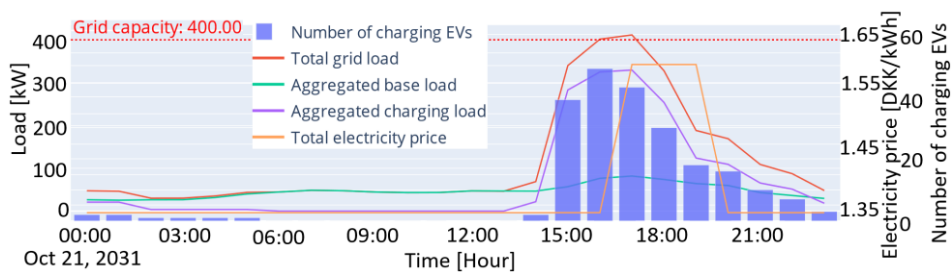

**Fig. 12.** Details of the day when the first overload occurs in the grid.

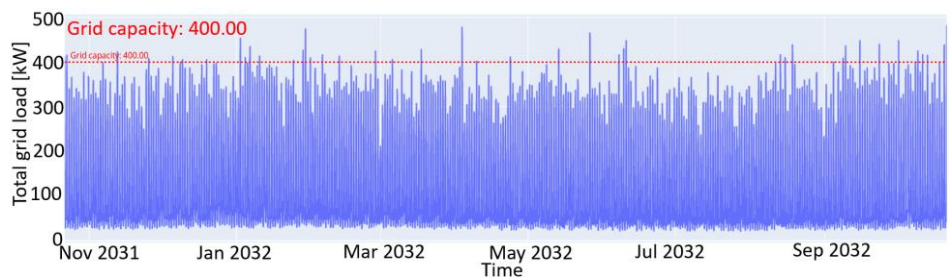

**Fig. 13.** Total grid loads the year after the first overload (October 21, 2031, to October 21, 2032).

By the end of 2030, 69 EVs were adopted. This resulted in an average annual emission in 2031 of 116.37 kg per EV. Fig. 14 shows the individual CO2-eq. emission from charging the EV, with the colors representing the different EV models. The individual emissions vary depending on the driving and charging behavior as evident from Fig. 14.

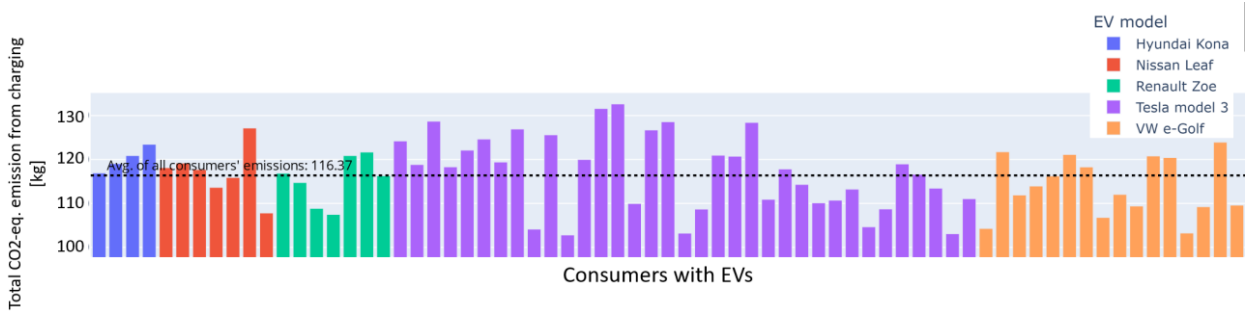

**Fig. 14.** The total annual CO2-eq. emissions from charging for 69 EVs in 2031.

### 7.2 Various Electric Vehicle Adoption Curves Scenario Results

The various EV adoption curve scenario considers the adoption of 775k and 1 million EVs in Denmark by 2030, based on the estimation and the political goal by [32], respectively. This scenario investigates when overloads can be expected with different EV adoptions and differentiates only by the EV adoption compared to the baseline. To evaluate the two EV adoption curves, the simulation follows the actual EV growth instead of the Poisson process used for the baseline scenario and avoids a situation where one curve adopts faster than another. The result (shown in Table 4) proves that the adoption curve with 1 million EVs by 2030 results in an eight-month earlier overload compared to the adoption curve with 775k EVs by 2030.

The average CO2-eq. emissions are close to the same as in the baseline scenario, which is no surprise since the emission levels are the same, but with fewer EVs adopted. For the 775k and 1 million EVs by 2030 experiments, the number of EVs at the end of 2030 is 30 and 42, respectively.

**Table 4.** Key results for electric vehicle adoption curve scenario.

| Experiment | Overload date | Number of overloads the year after the first overload | Days with overload | Total number of EVs at overload time | Avg. total CO2-eq. emissions in 2031 [kg] |
|---|---|---|---|---|---|
| 775k EVs by 2030 | October 26, 2033, 4 PM | 8 | 7 | 71 | 115.72 |
| 1 million EVs by 2030 | February 20, 2033, 6 PM | 15 | 10 | 77 | 114.21 |

Both scenario experiments (775k and 1 million by 2030) are adopting slower than the adoption estimated based on the historical data (1.3 million by 2030) used in the baseline scenario. The slower adoption results in around 16 and 24 months later overload. The result indicates that the DSO's planning for dealing with grid overloads should be proactive in that the overloads might happen much earlier than the estimation, especially due to the rapid adoption of other distributed energy resources in the distribution grids.

## 8 Discussion

It is no supervise, the results show that the adoption curve based on historical data and used in the other scenarios has a faster adoption (1.3 million EVs by 2030) than the 1 million by 2030 adoption curve. The current Danish regulations are following the suggestions to reach 775k EVs by 2030 because reaching 1 million EVs is too costly. However, the Danish authorities are still aiming for 1 million EVs by 2030 and are reconsidering new regulations in 2025 to reach the goal [33].

The reason for considering the curve determined based on historical data for the case study area (Strib) is due to the relation between income and EV adoption as shown in Fig. 15. Fig. 15 shows the EV population (left in the figure) and average personal income (right in the figure) for the municipalities in Denmark. The city of Strib (marked by the red circles) is part of one of the municipalities with a relatively high average income. Hence, a faster adoption in the city of Strib than the average adoption of the whole Danish population is expected.

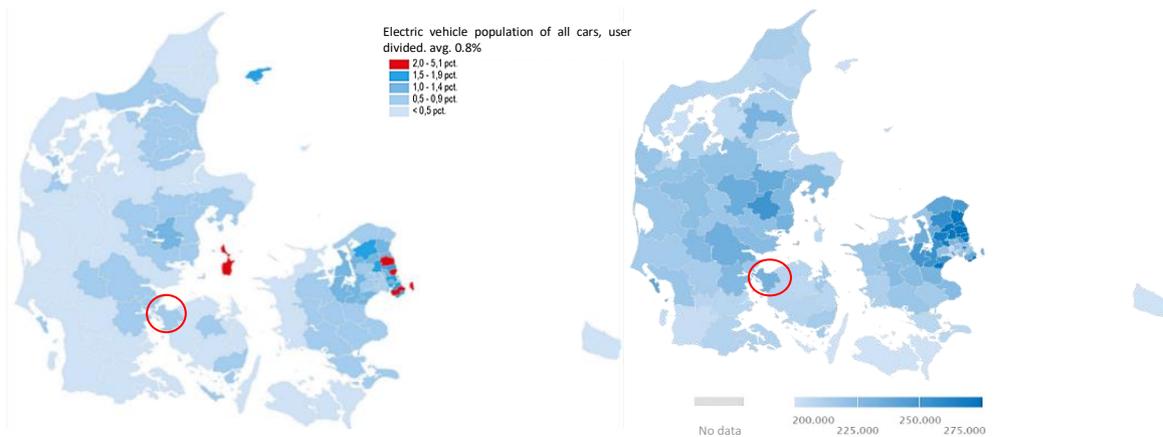

**Fig. 15.** Maps of Denmark illustrate the electric vehicle population [until August 2020] [34] (left) and the average personal income [DKK in 2017] in the municipalities [35] (right).

As mentioned at the beginning of this paper, the average CO2-eq. emissions from cars in Denmark were around 2.72 tons annually in 2019. The simulation results showed that for EVs in 2031, the average emission is around 115 kg per EV. At this time, the electricity mix consists of a high share of green production, resulting in low emissions from consuming electricity. With 775k EVs in 2030 out of 3.25 million would result in 6.821 million tons of CO2-eq. emission (calculation shown in equation 3). This does not meet the limit of 4.97 million tons necessary to satisfy the goal of a 30% reduction compared to 2005 emissions.

$$775{,}000 \cdot 115 \text{ kg} + (3{,}250{,}000 - 775{,}000) \cdot 2{,}720 \text{ kg} = 6.821 \text{ million tons} \quad (3)$$

For the scenarios with 1 million and 1.3 million EVs, this resulted in 6.235 and 5.456 million tons, respectively. This does also not satisfy the goal, as this requires at least 1.486 million EVs by 2030 to achieve a reduction of 30% of emissions compared to 2005. Alternatively, a larger reduction should be achieved in the other groups of the transport sector, since this calculation only considers passenger cars. As conventional cars also become more efficient, the exact number of EVs needed might be lower.

A faster adoption might help achieve the climate goal, however, this results in new challenges for the DSO. The results show that the current residential distribution grid can manage around 43-49 simultaneous charging EVs depending on the EV models and charging behavior. Each of these experiments uses different randomness (i.e., the random values generated in the simulation differ between scenarios) in the simulation due to the different adoption rates. However, with the conditions presented in the three scenarios, the overload is expected to occur between October 2031 and October 2033. The faster the EV adoption, the faster the grid experiences the first overload, with more frequent overloads following.

## 9 Conclusion

This paper investigates how different adoption curves of private electric vehicles (EVs) will impact the national climate goals with an example of Denmark's 2030 climate goals and the consequence to the electricity distribution grids. The results show that both considered EV adoptions (1 million and 775k EVs by 2030) will not satisfy the Danish obligation to the European Union's climate goal of reducing transport sector emissions by 30% by 2030. Furthermore, the results show that the current electricity distribution grids cannot handle the increasing load from EVs. With an EV adoption extrapolated from historical data and the traditional charging strategy (immediately start charging at arrival without any control), the overload will occur in 2031 with a 67% share of EVs in the grid. With a slower adoption, the first overload is experienced around 16 and 24 months later for the 1 million and 775k EVs by 2030 goal and estimation, respectively.

This paper enlightens uncertainties from two aspects: an unclear loading profile of distribution grids in the future due to EV adoption, and the indirect CO2-eq. emissions from charging EVs. As stated in the introduction section, the majority of the literature is based on static modeling or assumption-based simulations without considering the real EV adoption curves and varying CO2-eq. emissions from charging.

This paper applies multi-agent based simulation with two adoption curves, a common EV charging strategy (Traditional charging), various EV models, driving patterns, and CO2-eq. emission data to capture the full ecosystem dynamics during a long-term period from 2020 to 2032 with a high resolution (hourly). The research provides a clear load profile of distribution grids as well as the indirect CO2-eq. emissions from EV charging in the future due to EV adoption.

In this research, the financial aspects of EV charging under different price structures are not considered and are recommended for future research to be included e.g., the financial impact from hourly electricity prices with dynamic tariffs. Moreover, this paper mainly focuses on the impact of EV adoption curves on the electricity distribution grids, therefore, only Traditional EV charging strategy is considered in the simulation. However, there are various EV charging algorithms discussed in the literature, e.g., centralized and decentralized EV charging which future research should take into consideration.

## Acknowledgments

This research is part of the national projects of Flexible Energy Denmark FED funded by Innovation Fund Denmark (Case no.8090-00069B), and Digital Energy Hub funded by the Danish Industry Foundation.